# Costly Collaborations: The Impact of Scientific Fraud on Co-authors' Careers

Philippe Mongeon[a] and Vincent Larivière[b]

[a]École de bibliothéconomie et des sciences de l'information, Université de Montréal, Montreal H3C 3J7, Canada

[b]Centre interuniversitaire de recherche sur la science et la technologie, Université du Québec à Montréal, Montreal H3C 3P8, Canada

Corresponding author:

Philippe Mongeon

École de Bibliothéconomie et des Sciences de l'Information, Université de Montréal CP 6128, Station Centre-Ville, Montreal, Quebec, H3C 3J7 Canada

philippe.mongeon@umontreal.ca

1-514-343-5600

This work was supported by the Centre Interuniversitaire de Recherche sur la Science et la Technologie (CIRST) and the Canada Research Chair program.

Abstract

Over the last few years, several major scientific fraud cases have shocked the scientific community. The number of retractions each year has also increased tremendously, especially in the biomedical field, and scientific misconduct accounts for approximately more than half of those retractions. It is assumed that co-authors of retracted papers are affected by their colleagues' misconduct, and the aim of this study is to provide empirical evidence of the effect of retractions in biomedical research on co-authors' research careers. Using data from the Web of Science (WOS), we measured the productivity, impact and collaboration of 1,123 co-authors of 293 retracted articles for a period of five years before and after the retraction. We found clear evidence that collaborators do suffer consequences of their colleagues' misconduct, and that a retraction for fraud has higher consequences than a retraction for error. Our results also suggest that the extent of these consequences is closely linked with the ranking of co-authors on the retracted paper, being felt most strongly by first authors, followed by the last authors, while the impact is less important for middle authors.

## Introduction

Retractions of scientific articles have skyrocketed in the last decade, especially in the medical field (Zhang & Grieneisen, 2012). More than 500 articles were retracted in PubMed in 2012 and 2013, which is more than one paper every day and a twentyfold increase compared to the average of 25 retractions per year from the 1990s. This tremendous increase of retractions, and the fact that about half of them are due to scientific fraud (Steen, Casadevall, & Fang, 2013), has been fuelling many discussions around scientific misconduct in both the scientific community and the general public.

Previous research has mostly focused on the rise of retractions (Cokol, Ozbay, & Rodriguez-Esteban, 2008; Steen, 2011), its causes (Fang, Steen, & Casadevall, 2012; Steen et al., 2013) and the ongoing citations of retracted papers (Furman, Jensen, & Murray, 2012; A. Neale, Northrup, Dailey, Marks, & Abrams, 2007; A. V. Neale, Dailey, & Abrams, 2010; Pfeifer & Snodgrass, 1990). Others have investigated the prevalence of scientific fraud (Fanelli, 2009; Sovacool, 2008; Steen, 2011), its potential consequences for science in general and for the public (Steen, 2012) as well as potential ways to prevent, detect and act upon scientific fraud (Steneck, 2006). While a few studies have looked at the consequences of fraud for particular disciplines (Azoulay, Furman, Krieger, & Murray, 2012) and for research teams (Jin, Jones, Lu, & Uzzi, 2013), our study is the first to compare the pre- and post-retraction productivity, impact and collaboration practices of all individual co-authors of retracted papers.

Many researchers have seen their scientific careers end in disgrace after being found guilty of scientific fraud, but fraudulent researchers rarely work alone. The case of Dutch psychologist Diederik Stapel alone has cast a shadow over the work of over 30 of his co-authors. Given the growth of collaboration since the 1950s (Wuchty, Jones, & Uzzi, 2007), an increasing number of researchers are expected to eventually be affected by a co-author's misconduct. Retraction notices or investigations generally identify specific author(s) as responsible for the fraud, but it is assumed that, despite being declared free of responsibility for the misconduct, innocent co-authors still suffer some consequences of the fraud (Bonetta, 2006). However, no empirical evidence has been provided to support this assumption yet. This paper fills this gap by investigating the post-retraction productivity (number of articles published per year), impact (number of citations per paper), and collaboration practices (number of authors, institutions, and countries per article) of co-authors involved in cases of scientific misconduct in the biomedical field.

Robert Merton (1968) has described the scientific system as a stratified and hierarchical space in which scientists make contributions to the common stock of knowledge in exchange for different forms of recognition, or what Bourdieu (1986) calls symbolic capital. In turn, scientists use their symbolic capital to gain greater access to resources and increase the number or importance of their contributions, and consequently gain more symbolic capital and improve their position in the structure of science. This mechanism of accumulation of symbolic capital is accentuated by what Merton (1968) called the Matthew effect, which can be defined as the bestowing of greater recognition upon those who already have it, and the denial of recognition for those who do not. In the context of collaborative research, the symbolic capital obtained with a given contribution is shared between its co-authors, and one can expect that the negative capital obtained when a discovery is found to be fraudulent is also shared to some extent. If that is the case, this negative effect might be observed in at least one of three ways: 1) the researchers might publish less papers; 2) their publications might be less cited by their peers, and 3) they might be less prone to collaboration, either because they become more hesitant to collaborate, more selective in terms of people they are willing to work with, or because other scientists are less inclined to collaborate with them because of their link to a case of misconduct. Thus, this paper analyses the productivity, scientific impact and collaboration practices of co-authors; a significant decline of these indicators following the retraction of a paper would indicate the price to be paid for one's association with a publicly-known case of misconduct.

In order to evaluate the extent of these consequences, we looked at the career progression of 1,038 researchers in the biomedical field who contributed to a retracted paper between 1996 and 2006. For a period of five years before and five years after the retraction, we calculated the authors' number of publications, as well as their papers' characteristics, such as average relative citations and number of co-authors, institutions and countries. For comparison, we used a control group of 1,862 co-authors who were not involved in any known case of fraud. We examined the first, middle, and last authors of retracted papers separately since, in the biomedical field, the distribution of credit (and responsibility) amongst co-authors typically has a U-shape with the first author(s) generally having led the work, the last author(s) having supervised it, and middle author(s) being less involved (Pontille, 2004). Thus, one could argue that the consequences of scientific fraud for authors should be proportionate to their individual level of responsibility. We also looked separately at the authors of papers retracted for fraud and those retracted for error. While there is a general agreement that honest mistakes are normal in the course of science, and that they "must be seen not as sources of embarrassment or failure, but rather as opportunities for learning and improvement" (Nath, Marcus, & Druss, 2006), fraud is a serious deviation from the core values and the purpose of scientific research. Therefore, we expect that a retraction for fraud will have more impact on a researcher's career than a retraction for error.

This study is to our current knowledge the first to provide empirical evidence of the consequences that retractions in the biomedical field (most importantly those that occur in cases of scientific fraud) have on the careers of the co-authors who are not formally identified as responsible for the fraud or mistake. It is also the first study that provides data about the subsequent research output of the retracted papers' authors. This focus on the biomedical field is both a necessity and a limit, the former because the majority of retracted papers are in this field with only a few retractions occurring in other fields, and the latter because it restricts the generalizability of our findings to the biomedical field.

## Data and methods

**Retracted papers**

To obtain our sample of retracted papers, we searched PubMed for all retraction notices (publication type "Retraction of publication") and all retracted papers (publication type "Retracted publication"), finding a total of 2,451 retractions and 2,299 retracted papers. We paired retracted papers with their corresponding retraction notice in order to find the year of retraction and the delay of retraction (i.e., the number of years between publication and retraction) of all retracted papers. We then found these retracted papers in a bibliometric version of Thomson Reuter's Web of Science (WoS), provided by the Observatoire des Sciences et des Technologies. It is a relational database which allows the linking of any variable to another, constructed from the XML source data provided by Thomson Reuters, and stored on a Microsoft SQL server. We limited our search to papers in English published in biomedical and clinical medicine journals. We kept only papers which were retracted between 1996 and 2006 inclusively, in order to obtain a sufficient time window for the assessment of the co-authors' productivity, impact and collaboration before and after the retraction. This provided us with a sample of 443 retracted papers.

In an analysis of retracted papers found in PubMed, Azoulay and collaborators (2012) classified retracted papers according to the cause of retraction. They did so using the information found on the retraction notice as well as any other information found on the web. We used their data to categorize the papers in our sample according to the cause of retraction: fraud (including data fabrication or falsification and plagiarism) (N = 179) and error (N = 114). Papers retracted for other reasons (N = 150) were not included in our analysis.

**Authors of retracted papers**

We listed all authors of the papers retracted for fraud or error. After author name disambiguation, which was done by looking at each retracted papers of authors with two or more retractions, the list contained 1,098 distinct authors. Then, for each of these authors, we searched the Web of Science for all papers published within five years before and five years after the retraction. For authors who retracted multiple papers on different years, we searched for papers published within five years before the first retraction and five years after the last. We used a five year interval in order to gather a sufficient number of publications for each researcher and to ensure that the changes observed were not simply due to long term trends or normal short term variations.

After author name disambiguation, the resulting sample was a total of 15,333 distinct articles, published between 1991 and 2011. The author name disambiguation was done manually using the name and initials of authors, their affiliations, the discipline of the journal, and keywords in the article titles. When this information did not allow us to distinguish homonyms, we looked at the article itself, or searched the web for the curriculum vitae of the researcher.

**Time to retraction**

To regroup articles that were published in a specific period relative to the retraction year (e.g. five years prior to retraction), it was necessary to convert the publication year into another variable that we call "time to retraction" (T). The value of T ranges from -5 to 5, with 0 being the year of retraction. In cases of multiple retractions on multiple years, T = 0 for all articles published between the year of the first and of the last retraction, inclusively. In most parts of our analysis we divided publications in two groups: pre-retraction and post-retraction. The first group includes papers published between T = -5 and T = -1, and the second one includes articles published between T = 1 and T = 5.

**Fraudulent and innocent researchers**

We used data from Azoulay and collaborators (2012) as well as the retraction notices and web searches to identify authors responsible for the fraud or other retraction cause. We were able to identify the responsible authors for most cases of fraud (82 authors, responsible for 159 of the 179 fraud cases), while there was rarely a responsible author identified in cases of error (3 authors, responsible for 5 of the 114 error cases.). No responsible authors could be identified for 20 cases of fraud (4 data fabrication or falsification cases and 16 plagiarism cases). Since it is impossible in those cases to distinguish the fraudulent authors from the innocent ones, all authors of these 20 papers (N = 66) were removed from our sample, in order to ensure that it contains only innocent researchers.

We also divided co-authors into three exclusive groups, according to their rank in the author list of the retracted paper. In cases of authors with multiple retractions, they were assigned to the group "first authors" if they were first author of at least one retracted paper. The "last authors" group contains authors who were listed as last author on at least one retracted paper, and who were not listed as first author on any retracted paper. The "middle authors" group contains authors who were assigned to neither of the "first authors" and "last authors" group. For single authored papers, the author was considered first author, while in cases of papers with two authors, the second one was assigned to the "last author" group.

**Control group**

For each of the articles retracted between 1996 and 2006 in our original list (i.e. including retracted papers for other reasons than fraud and error), we randomly selected a non–retracted article with the same number of authors, published in the same issue of the same journal. This provided us with a list of 1,862 authors, for which we retrieved all publications over the five years prior to and after the retraction of their corresponding retracted paper, for a total of 55,036 papers. The authors of the control group were also categorized according to their rank on the article published in the same journal as the retracted paper.

**Final sample**

The final sample of authors, including the control group, is shown in table 1

*Table 1*

*Sample of authors.*

| Rank | Fraud | Error | Control | Total |
|---|---|---|---|---|
| First authors | 45 | 108 | 411 | 564 |
| Middle authors | 346 | 366 | 1,046 | 1,758 |
| Last authors | 77 | 102 | 405 | 584 |
| Total | 468 | 576 | 1,862 | 2,906 |

Authors who had no publication in either the pre- or the post-retraction period were removed from the sample used to assess the co-authors' scientific impact (see figure 2 in results section) and collaboration practices (see figure 3 in results section). The sub-sample used for these analyses is shown in table 2.

*Table 2*

*Sub-sample of authors with at least one publication in the pre- and post-retraction periods.*

| Rank | Fraud | Error | Control | Total |
| --- | --- | --- | --- | --- |
| First authors | 28 | 83 | 354 | 465 |
| Middle authors | 253 | 276 | 860 | 1,389 |
| Last authors | 64 | 89 | 382 | 535 |
| Total | 345 | 448 | 1596 | 2,389 |

**Bibliometric indicators**

To measure the productivity of co-authors, we used the number of papers published per year. The number of papers was normalized at the individual level by dividing the value for a given year by its average over the period of five years prior to five years after the retraction. We call the resulting number the individual relative productivity (IRP) of the researcher. This individual-level normalization allows comparison of this indicator between researchers. To measure the scientific impact of co-authors, we used the average number of citations received by their papers. The number of citations for each paper was normalized at the discipline level, by dividing the number of citations of a paper by the average number of citations received by all papers published in the same field and in the same year, the field being determined by the journal in which the paper was published, and the journal's discipline being determined by the National Science Foundation journal classification. The resulting indicator is called average relative citations (ARC). Finally, we used the number of authors, institutions and countries listed on co-authors' publications, also normalized at the discipline level, as an indicator of their collaboration practices.

**Statistical tests**

To assess the significance of differences observed between groups, we used the Mann-Whitney U-test. The Mann-Whitney U-test was preferred to a t-test because it is more robust and because of the non-parametric nature of the compared distributions. The null hypothesis ($H_0$) of the Mann-Whitney U-test is that the compared groups have the same median, and the alternative hypothesis ($H_1$) is that the medians are not equal. A statistically significant difference ($P < 0.1$) means that there is less than 10% probability that the differences observed are due to chance and that the medians are, in fact, the same. Therefore, when we obtained a value of P smaller than 0.1, we rejected the null hypothesis and accepted the alternative hypothesis.

## Results

**Fraudulent authors**

Before looking at the innocent co-authors, we looked at the 79 co-authors who were officially identified as responsible for the fraud (N = 79). Of those co-authors, 45 had no publication in the five years following the retraction, presumably because they left the scientific field, while the median IRP of the 34 remaining authors decreased by 64.6%. These results confirm that the discovery of a fraud will strongly affect the career of fraudulent researchers, putting an end to it in most cases. Those 79 co-authors (as well as the 3 authors who were identified as responsible for errors) were excluded from the sample for the other analysis, which focused on the innocent collaborators.

**Innocent collaborators**

Most striking is the number of co-authors for whom we found no publications in the five years following retractions. As shown in Table 3, this is the case for 27.6% and 20.3% of co-authors of papers retracted for fraud and errors, respectively. There were much fewer (12.1%) co-authors in the control group with no publications in the five year post-retraction period.

*Table 3*

*Proportion (%) of authors with no publications in the five years following retraction.*

|  | Fraud | | Error | | Control | |
| --- | --- | --- | --- | --- | --- | --- |
| Rank | N | % | N | % | N | % |
| First authors | 17 | 39.1 | 24 | 22.2 | 63 | 11.3 |
| Middle authors | 96 | 27.7 | 81 | 22.1 | 266 | 15.2 |
| Last authors | 16 | 20.8 | 12 | 11.8 | 29 | 5.0 |
| Total | 129 | 27.6 | 117 | 20.3 | 358 | 12.1 |

Thus, unsurprisingly, figure 1 shows that the median individual relative productivity (IRP) of co-authors of retracted papers quickly drops in the years following a retraction. Figure 1 also shows that the extent of this drop of productivity depends on the reason for retraction (productivity losses are more important after a retraction for fraud than after a retraction for error) and the position of the co-author on the retracted article's byline (first authors' IRP decreases more abruptly than middle and last authors). Figure 1 also shows that IRP of authors in the control group evolves differently for first, middle and last authors. While the median IRP is quite stable during the entire 11 year period for last authors, we see a higher variation for first and middle authors. For first authors, there is a sharp increase in the pre-retraction period, followed by a slight decrease. As for middle authors, the median IRP also raises in the pre-retraction, with a similar decrease in the post-retraction period. This variation is most likely due to the fact that last authors are usually senior researchers with stable careers, while first and middle authors can be transient authors who may not pursue a scientific career.

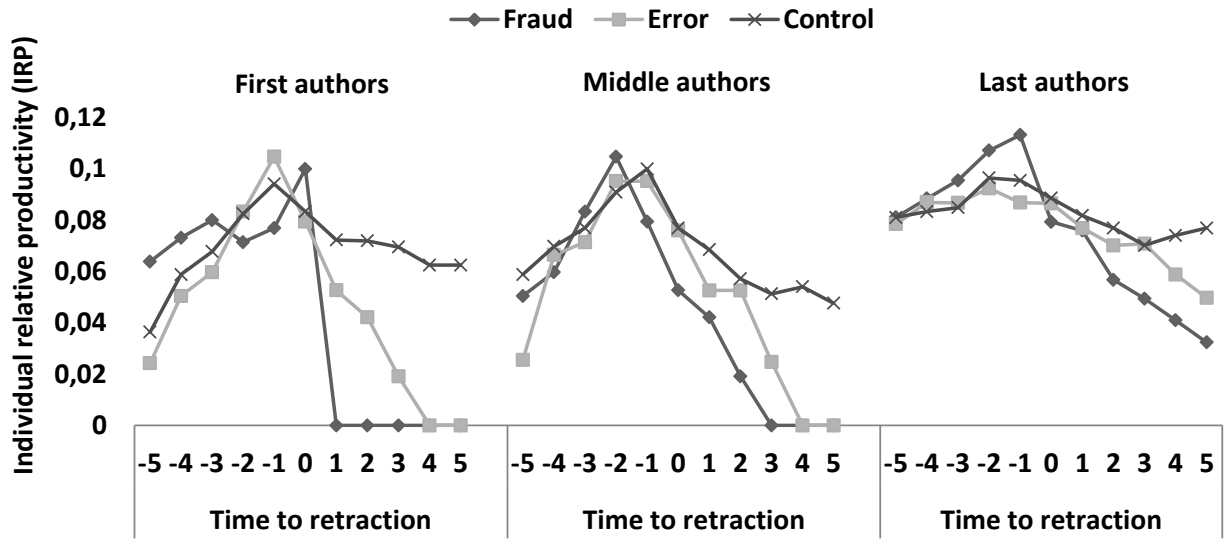

*Figure 1. Median individual relative productivity (IRP) from five years prior to five years after the retraction.*

The median proportion of papers published before and after the retraction (Table 4) shows that co-authors suffer a significant decrease in IRP, no matter the reason of retraction, with the exception of last authors of papers retracted for error, for whom the decrease in productivity doesn't prove to be statistically significant. Fraud cases, but not errors, have a significant impact on middle authors' IRP. However, middle authors of the control group also show an important decrease in IRP. Therefore, our results suggest that the retraction has less impact on the subsequent number of publications of middle authors, compared to first and last authors. Thus, the extent of consequences felt by co-authors seems to be distributed in a way that is similar to the distribution of the credit received by the authors, with the first and last authors being more affected by the scientific fraud of their co-authors than middle authors.

*Table 4*

*Difference between the pre- and post-retraction median individual relative productivity (IRP).*

| Rank | Group | IRP variation (%) | P-Value[*] |
|---|---|---|---|
| First authors | Fraud | -65.2 | .000[**] |
|  | Error | -35.5 | .006[**] |
|  | Control | 0 | - |
| Middle authors | Fraud | -50.0 | .000[**] |
|  | Error | -44.2 | .000[**] |
|  | Control | -25.0 | - |
| Last authors | Fraud | -46.7 | .000[**] |
|  | Error | -23.5 | .102 |
|  | Control | -17.4 | - |
| All authors | Fraud | -50.0 | .000[**] |
|  | Error | -39.2 | .000[**] |
|  | Control | -18.3 | - |

*Note:* Co-authors who only published articles at T = 0 are not included here. This is the case of 8, 10 and 14 co-authors in the fraud, error and control groups, respectively.
[*]The P-value results from a comparison of the group of authors with the control group using the Mann-Whitney U-test.
[**]The decrease in publication for these groups of co-authors is statistically significant ($P < 0.01$).

In assessing the effect of retractions on co-authors' scientific impact, we had to exclude from our sample the researchers for whom we did not find any publications in either of the pre- or post-retraction periods. This reduced our sample considerably (see table 2 in the methods section) since, as we saw earlier, many researchers had no publications after the retraction. Figure 2 shows, for each group of co-authors, the median decrease in average relative citations (x-axis) and individual relative publications (y-axis) of authors from the sub-sample. Groups with an increase of median IRP and median ARC would be located in the first quadrant of the graph (top right), while groups with a decrease for both indicators would be located in the third quadrant (bottom left).

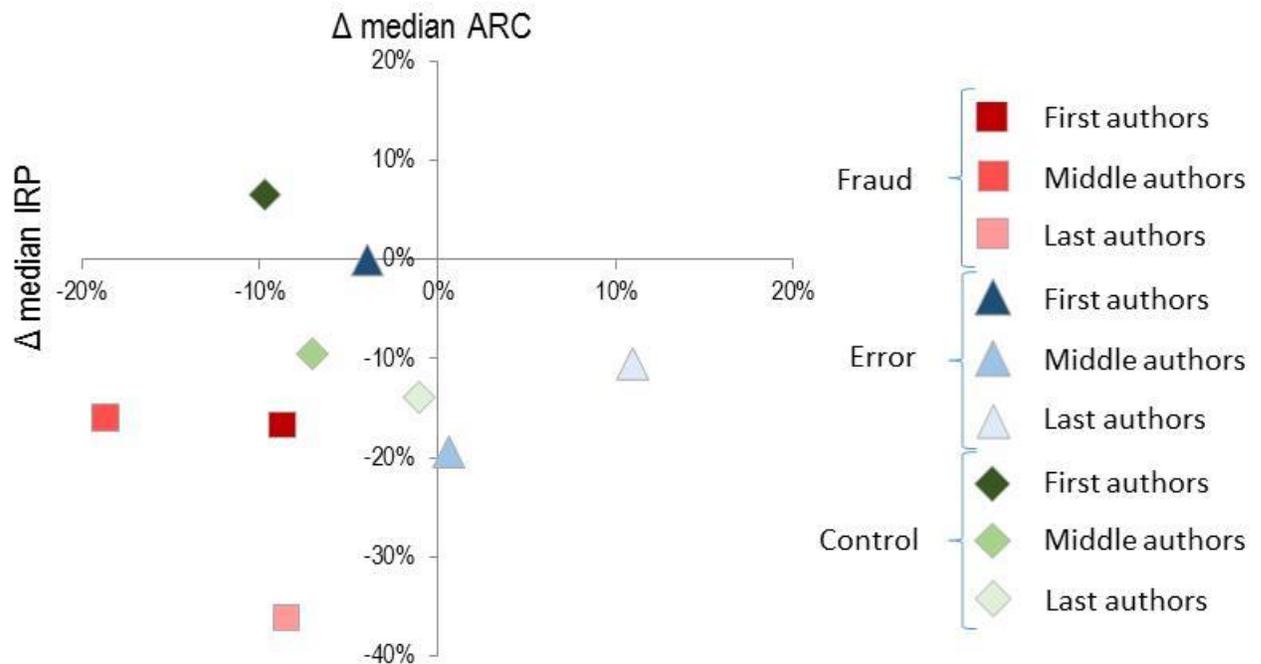

*Figure 2. Difference in the pre- and post-retraction median individual relative productivity (IRP) and median average relative citations (ARC).*

Figure 2 clearly illustrates the distinction between co-authors who retracted a paper for fraud, those who retracted a paper for error, and those belonging to the control group. As one might expect, the median productivity and impact of authors belonging to the fraud group are situated at the lower left of the graph, indicating that both these indicators are negatively affected by the fraud. Also as expected, the medians for the control group are closer to the middle of the graph, indicating a relative stability over the pre- and post-retraction periods. Interestingly, authors who retracted an article for error seem to see their scientific impact decrease slightly less than authors of the control group, and even increase in the case of middle and last authors. In terms of scientific impact alone, most of the differences observed are not statistically significant ($P > 0.05$), except for middle authors of articles retracted for error, for whom the median impact has significantly increased ($P < 0.05$) (see table 5). This limitation in our study can be partly explained by the fact that the many researchers who didn't publish after the retraction were excluded from this part of the analysis. However, the results shown in table 5 confirm that, when dividing authors only by cause of retraction (and not by rank on the paper), fraud has a significant negative impact on citations at a 90% confidence interval ($P < 0.1$) while, on the other hand, errors have a positive impact on citations at a 99% confidence interval ($P < 0.01$).

*Table 5*

*Difference between the medians of the pre- and post-retraction average relative citations (ARC)*

| Rank | Group | ARC variation (%) | P-Value* |
|---|---|---|---|
| First authors | Fraud | -8.7 | .986 |
| | Error | -4.0 | .274 |
| | Control | -9.7 | - |
| Middle authors | Fraud | -18.7 | .092 |
| | Error | 0.6 | .013*** |
| | Control | -7.1 | - |
| Last authors | Fraud | -8.5 | .524 |
| | Error | 11.0 | .108 |
| | Control | -1.0 | - |
| All authors | Fraud | -17.6 | .056**** |
| | Error | 2.0 | .003** |
| | Control | -7.5 | - |

*The P-value results from a comparison of the group of authors with the control group using the Mann-Whitney U-test.
**The difference in average relative citations for these groups of co-authors is statistically significant ($P < 0.01$).
***The difference in average relative citations for these groups of co-authors is statistically significant ($P < 0.05$).
****The difference in average relative citations for these groups of co-authors is statistically significant ($P < 0.1$).

This may be linked to a previous finding by Lu, Jin, Uzzi, & Jones (2013), who showed that self-reported retractions (most likely errors) led to an increase in citations for the authors' previous work. Our results would then suggest that this might also be the case for the authors' ulterior work. Also, while retractions have a relatively small impact on the field-normalized average number of citations received by further papers, there is nonetheless a more important decrease in the overall impact, as measured by the total number of citations received, due to the decrease observed in publications (see Figure 1).

**Collaboration**

Finally, we assessed the impact of retraction on co-authors' level of collaboration. We found no observable difference in the number of authors, institutions or countries per paper normalized at the discipline level in the pre- and post-retraction periods. For example, Figure 3 shows the level of inter-institutional collaboration over the whole period, which appears to be similar for all groups. Thus, retractions do not appear to have any effect on the collaboration practices of co-authors.

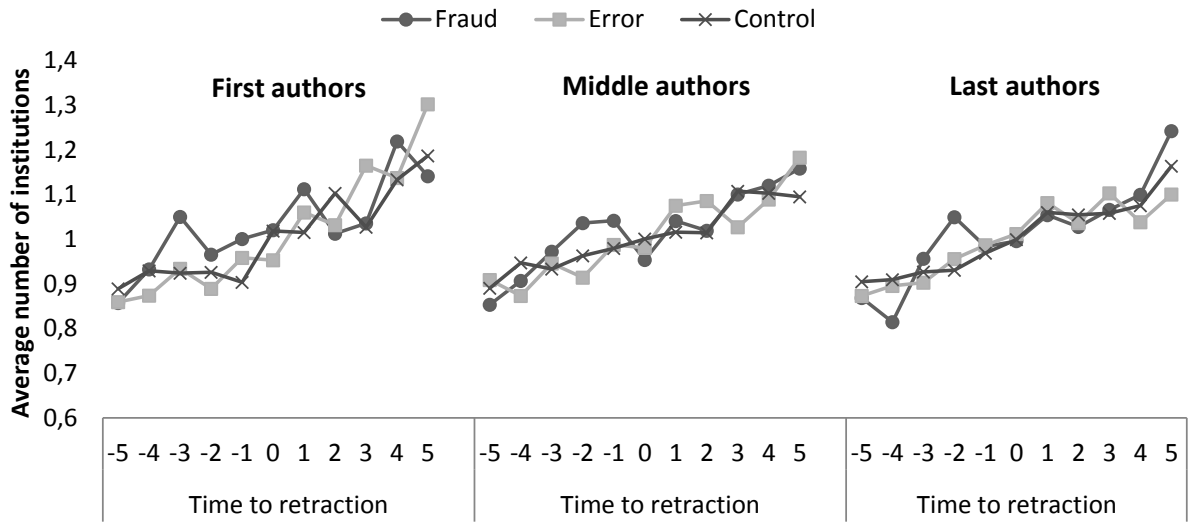

*Figure 3. Average number of institutions per paper normalized at the discipline level from five years prior to five years after the retraction.*

## Discussion

We found clear evidence that collaborators do suffer consequences of their colleagues' misconduct, and that fraud has more consequences than errors. Our results also suggest that the extent of these consequences is closely linked with the ranking of co-authors on the retracted paper, being felt most strongly by first authors followed by last authors, while the impact is less important for middle authors. Notably, the important difference in the impact of a retraction for first and last authors is most likely due to the fact that, while the former are often junior researchers with a more precarious professional status, the latter are generally well-established senior researchers whose position in the social structure of science may help in overcoming the shock of the retraction. Nevertheless, our findings are consistent with the idea that greater involvement in the research process leads not only to greater credit, but also to greater responsibility (Birnholtz, 2006), and therefore to greater consequences in cases of fraud. In the context of collaborative work, the symbolic capital acquired through the publication of a discovery is shared, though unequally, between all co-authors. Thus, it might be said that retracting an article found to be fraudulent brings to its co-authors a negative form of symbolic capital (i.e., blame instead of recognition), which is also shared unequally (the fraudulent author most likely getting the largest share). Thus, linking our findings to the theory of cumulative advantage and disadvantage (Merton 1968) suggests that the retraction of a fraudulent paper might initiate a negative trend that can be difficult to reverse for some researchers, leading many of them, mostly the least experienced ones, to abandon the field entirely.

As mentioned in the introduction, we expected that the consequences of retracting an erroneous paper would be minimal. However, our results show that errors do have consequences (though not as important as in cases of fraud) for collaborators in terms of productivity. These results might be partly explained by the fact that retractions occur generally in cases of major errors that invalidate the findings as a whole, while minor errors lead most likely to corrections. However, errors seem to have a positive effect in terms of citations. More studies might be necessary to gain a better understanding of this phenomenon.

One limitation of our study is that, because of the high number of researchers who stopped publishing altogether after the retraction (mostly in cases of fraud), we are left with a sample that is quite small for our comparison of the pre- and post-retraction impact and collaboration. In this regard, a future study repeating our methodology could benefit from the much higher number of retractions that have occurred in the biomedical field in the last few years. Another limitation is that the methodology used doesn't allow us to control for factors other than the retraction that could affect researchers' productivity, scientific impact and collaboration. However, our use of a large control group and the significance measures obtained with the statistical tests reinforces the confidence in the validity of our results.

Being limited to the biomedical field, this study does not enable us to generalize our findings to other disciplines. Thus, further research could assess the consequences of retractions for co-authors in other disciplines. However, retractions are not as prevalent in other disciplines as they are in biomedical research, which might make it more difficult to gather enough retractions to do an analysis like the one presented here. Future work could also look at how the effect of retractions varies depending on socio-cultural or institutional factors. This is also something that might be made possible by the larger number of retractions in recent years.

## Conclusion

Our results show that scientific fraud in the biomedical field is not only harmful to science as a whole and, on an individual level, to the fraudulent scientist, but also to the innocent scientists whose only fault might have been choosing to work with the wrong colleague. Indeed, many of the co-authors of fraudulent papers included in our study have published fewer papers in the years following the retraction of a fraudulent paper. To a lesser extent, the average number of citations received by their papers has also decreased following the retraction. However, the retraction did not seem to affect their collaboration practices. These changes in productivity and impact can be linked to the co-authors' association with a case of fraud, since neither of the co-authors of our control group or those who retracted papers because of errors, show similar trends. The latter even seem to see the average number of citations of their papers increase after the retraction.

The effect of having participated in a case of scientific fraud goes way beyond a decrease in papers or loss in scientific impact for the fraudulent authors and their collaborators. Some consequences can be psychological (e.g., scientists losing trust in science, colleagues and institutions) or a waste of research efforts and funds. The case of Hendrik Schön in physics provides a good example of this waste of efforts: he forged 'ground-breaking' results that many other researchers around the globe were eager to reproduce and build upon. Much time and many funds were wasted in those inevitably unsuccessful attempts, and the discovery of the fraud led a few discouraged scientists (mostly PhD and postdoctoral students) who had been working on these projects to abandon the idea of pursuing a career in research (Reich, 2009). Moreover, the cases of fraud that are discovered almost every day are most likely only the tip of the iceberg: a meta-analysis of surveys reported that 2% of scientists admitted to having falsified or fabricated data, while 14% said they knew someone who did (Fanelli, 2009). In the United States, allegations of fraud received by the U.S. Office of Research Integrity (ORI) have increased to a point where only a small proportion can actually be investigated ("Seven days: 26 April–2 May 2013," 2013). It is therefore likely that the number of cases will keep rising and that more and more collaborators will see their careers compromised. Finally, by being able to get a firsthand look at the work of their collaborators before it is submitted for publication, co-authors are the first link of the peer-review chain. Considering the potential consequences that co-authoring a fraudulent paper can have on their careers, it is in their best interest to take this role seriously.

COSTLY COLLABORATIONS                                                                 21

COSTLY COLLABORATIONS                                                                                        22